
\magnification 1200
\centerline{\bf RENORMALIZABILITY OF SEMIQUANTIZED FIELDS }
\vskip1cm
\centerline{ L.L. Salcedo\footnote{*}{e-mail: salcedo@ugr.es}}
\vskip0.5cm
\centerline { Departamento de F\'{\i}sica Moderna, Universidad de Granada}
\centerline {E-18071 Granada, Spain }
\vskip1cm

\centerline{\bf ABSTRACT}
\vskip 18pt
\vbox{
\baselineskip 12pt
\narrower{
A definition is given, in the framework of stochastic quantization,
for the dynamics of a system composed of classical and
quantum degrees of freedom mutually interacting.
It is found that the theory breaks reflection positivity, and hence
it is unphysical.
The Feynman rules for the Euclidean vacuum expectation
values are derived and the perturbative renormalizability of the theory
is analyzed.
Contrary to the naive expectation, the semiquantized theory turns out
to be less renormalizable, in general, than the corresponding completely
quantized theory.}}
\vskip1cm
\vfill
\hbox{UG-DFM-28/94\hfill}
\eject

\null\medskip
{\bf 1. Introduction }\par
\medskip
The study of the dynamics of systems mixing
quantum and classical degrees of freedom, has
attracted some interest in the past related to the interplay
of gravity and quantization.
Indeed Einstein's field theory is notoriously difficult to quantize
due to its lack of renormalizability [1,2].
This problem has been dealt with in the literature by considering
quantum gravity as an effective theory [3,4].

Alternatively, under the generic name of semiclassical gravity,
there has been a number of proposals to couple quantum matter
fields to a classical (or quasi-classical) gravitational field.
One of the hopes in this kind of approach was that it would reduce the
ultraviolet divergences which plague quantum gravity.
Classical field here does not mean background. The classical field is
dynamical, yet it is not quantized. The precise meaning of this
varies in the different approaches found in the literature.
In the older and more extended formulation [5-7],
the classical field
couples to the expectation value of the energy-momentum tensor of the
quantum matter field, thus it behaves deterministically.
In this line, corrections to the expectation value to include
quantum fluctuations has been considered in [8] using the influence
functional formalism.

More recently an interesting proposal was made in [9].
The mixed system is described in the canonical formalism
by a density matrix containing both classical and quantum
variables. The density matrix evolves according to a linear equation of
motion which reduces to the commutator or to the Poisson bracket
with the Hamiltonian in the limits of purely quantum or purely
classical cases. Furthermore the evolution preserves the hermiticity
and the trace of the density matrix, and it is invariant under classical
canonical transformations and quantum unitary transformations.
However, the requirement of positivity of the density matrix
cannot be maintained in general. A similar approach is that of ref. [10]
although this is formulated directly in
terms of the classical and quantum variables evolving in Heisenberg
picture and also differs in details from [9] (see also [10a,10b]).
In these approaches the
classical field does not evolve deterministically, rather it inherits
fluctuations from its coupling to the quantum variables.

In this work we shall not consider gravitation specifically. In fact, we
regard any semiclassical treatment not as fundamental but at best as an
approximation to the fully quantum-mechanical theory, including gravitation.
Rather, we study the dynamics of the mixed classical-quantum
system by itself. This is done by giving a new definition of this kind
of systems which allows their field theoretical aspects to be emphasized.
In particular we consider renormalizability, which is
precisely the original motivation which gave rise to the
studies mentioned above. The stochastic quantization formalism
is used only because our definition is very natural within it.
It is an open problem to establish the translation of our prescription
to the canonical formalism, and
also its relation to the definition of [9]. Similarly to that
reference, we also find a positivity problem in our formulation (cf.
section 3).

\medskip
{\bf 2. Semiquantized dynamics}\par
\medskip
For simplicity, we restrict ourselves to the case of real bosonic fields.
In order to avoid mathematical subtleties, present both in stochastic
and path integral quantizations, we shall consider a system described by
a strictly finite number of degrees of freedom, as it is the case of
field theories regularized in a finite lattice [14], and the
Euclidean formulation of the theory. Any given configuration of the
system is therefore described by the set of real variables
$\phi_i,\ i=1,\dots,N$, where the discrete label $i$ contains all
the coordinates of the fields including (discrete) spacetime,
spin, flavor, type of particle and so on. To recover a true
mechanical system (with continuous time coordinate) or
a true field theory, one must take the infrared and ultraviolet
limits. This can spoil some of the arguments valid in the finite case.
To determine which of the assertions survive as the number
of degrees of freedom diverge is not a trivial matter in general and each
case requires a separate study. Nevertheless, we shall make some
comments below about the important issue of the
renormalizability of the theory. Moreover, we shall assume that the
Euclidean action $S(\phi)$ is real and that it is a polynomial
strictly increasing at infinity. This is sufficient
to guarantee that the action is mathematically well-behaved. In
particular, it will be bounded from below. A further physical requirement
is that the action must be reflection positive, hence defining a
positive definite measure for observables in Minkowskian space.

The dynamical problem is considered solved if
the vacuum expectation values of arbitrary polynomial observables
${\cal O(\phi)}$ are known. These expectation values have the form
$$\langle {\cal O}(\phi)\rangle = \int
{\cal O}(\phi) P(\phi) [d\phi]\eqno(2.1)$$
where $[d\phi]$ is the Lebesge measure over R$^N$,
and $P(\phi)$ is a real and positive probability density, normalized
to unity, which in general can be a distribution.
This approach unifies the various dynamics to be
considered below, namely, classical, quantized or semiquantized.
The particular distribution function $P(\phi)$
distinguishes the various dynamics and theories (actions).
In the quantum theory $P(\phi)$ is given by the Boltzmann weight [14-16],
$$P_Q(\phi)=e^{-S(\phi)} \eqno(2.2)$$
$P_Q(\phi)$ will be
normalized by adding a suitable constant term to the action $S(\phi)$.
On the other hand, in the classical dynamics, $P(\phi)$ is a Dirac
delta distribution localized at the minimum of the action.

As we said in the introduction, there is a natural definition of
the semiquantized dynamics within the stochastic quantization approach [13].
For convenience in the presentation,
we sketch here the main ideas underlying this approach.
Excellent reviews on this subject can be found in [11,12].
In the stochastic quantization formulation, the field configuration describes
a continuous random walk in configuration space, $\phi_i(t)$,
where $t$ denotes the simulation or Langevin time,
not to be confused with the physical time. The equation governing the random
walk is suitably chosen so that $P(\phi)$ is obtained as the stable
stationary probability distribution. This implies that vacuum expectation
values coincide, for arbitrary initial conditions,
with a simultaneous stochastic and temporal average
$$\langle {\cal O}(\phi)\rangle = \lim_{T\to+\infty}{1\over T}
\int_0^T \langle\!\langle{\cal O}(\phi(t))\rangle\!\rangle dt
\eqno(2.3)$$
Here $\langle\!\langle\ \ \ \rangle\!\rangle$ is the average over the
random noise.

To reproduce the probability distribution of the quantum dynamics, $P_Q(\phi)$,
the field configuration is let to evolve
according to the following Langevin equation:
$$\partial_t\phi_i(t)=-\partial_i S(\phi(t)) +\eta_i(t)\,,
\qquad\quad i=1,2,\dots \eqno(2.4)$$
Here $\eta_i(t)$ are independent stochastic Gaussian variables normalized to
$\langle\!\langle\eta_i(t)\eta_j(t')\rangle\!\rangle =
2\delta_{ij}\delta(t-t')$. The Langevin equation is a stochastic differential
equation which is to be understood in It\^o's sense, that is,
as the limit of the Markovian process [16]
$$\phi_{i,n+1}=\phi_{i,n} - \partial_i S(\phi_{i,n})\epsilon
+\eta_{i,n}\sqrt{2\epsilon} \eqno(2.5)$$
where $\eta_{i,n}$ are independent normal Gaussian variables and the limit
is taken as $\epsilon\to 0^+$ keeping fixed the Langevin time. Note
that the $\phi_i(t)$ are not operators but stochastic c-number
variables.

The stochastic term $\eta_i(t)$ introduces the quantum fluctuations which
distinguishes the quantum dynamics from the classical one.
This is more clearly seen
by introducing the standard bookkeeping positive parameter $\hbar$, i.e.,
by rescaling $S\to S/\hbar$. For convenience we also rescale the
Langevin time $t\to\hbar t$ (and thus $\eta_i\to\sqrt{\hbar}\eta_i$),
so that the Langevin equation becomes
$$\partial_t\phi_i(t)=-\partial_i S(\phi(t)) + \sqrt{\hbar}\eta_i(t)\,,
\qquad\quad i=1,2,\dots \eqno(2.6)$$
In the so-called classical limit, $\hbar\to 0^+$,
the quantum fluctuations are switched off, this
equation becomes deterministic and the equilibrium is attained when
$\phi_i$ is a stable solution of the classical equations of motion
$$0=\partial_i S(\phi), \eqno(2.7)$$
that is, when $\phi_i$ is the classical vacuum of the theory, or more
generally, one of the minima of the Euclidean action. This
minimum will vary in the presence of external currents added to the
action, hence allowing to pick up classical excited states as well.
This point will be further taken in Section 6, when we discuss the
effective action.

Let us note that the related Minkowskian problem formally satisfies
the Wick rotated equations. Such equations can directly be used in
perturbation theory or in simple cases where an exact analytical
solution is available (see Section 3 below). In general they can
not be solved directly by the Langevin algorithm since the
Boltzmann weight is complex in this case.
They can be solved indirectly by using
the so-called complex Langevin algorithm, but it does not always work
properly even for well-behaved actions [17-19].
Another remark is that at any moment,
we shall be able to discuss the classical limit of
any of the formulae below by following the same procedure of introducing
the parameter $\hbar$.

To show that indeed $P_Q(\phi)$ is the equilibrium distribution of the
Langevin equation, consider the instantaneous probability density
$P(\phi,t)=\langle\!\langle\delta(\phi-\phi(t))\rangle\!\rangle$, which
depends on the initial conditions. It satisfies
the following Fokker-Planck equation, as it is readily shown [20]:
$$\partial_t P(\phi,t) =\partial_i((\partial_i S(\phi))P(\phi,t)
+\partial_i P(\phi,t)) \eqno(2.8)$$
It is immediate to check that $P_Q(\phi)$ is a stationary solution.
Furthermore, it is the unique stable solution.
Indeed, let $\Psi(\phi,t)= \exp({1\over 2}S(\phi)) P(\phi,t)$; this
quantity satisfies the following Euclidean Schr\"odinger-like equation [20a]
$$\eqalign{
-\partial_t\Psi(\phi,t) &= H_{\rm FP}\Psi(\phi,t) \cr
H_{\rm FP} &= -\partial_i ^2 +V(\phi) \cr
V(\phi) &= {1\over 4}(\partial_i S)^2-{1\over 2}\partial_i^2 S \cr
} \eqno(2.9)$$
The potential $V(\phi)$ is also polynomial and strictly increasing at
infinity, hence it has a unique ground state, namely, $\Psi_0(\phi)=
\exp(-{1\over 2}S(\phi))$, with zero energy. All the other eigenfunctions
have strictly positive eigenvalues and hence $P_Q(\phi)$ is a stable
fixed point of the Fokker-Planck equation. That $H_{\rm FP}$ is
semidefinite positive can also be seen by rewriting it in the form
$$H_{\rm FP} = Q_i^\dagger Q_i\,,\qquad
Q_i = \partial_i + {1\over 2}\partial_i S(\phi) \eqno(2.10)$$
In passing, it has been noted that $H_{\rm FP}\ge 0$ even if the
action is unbounded from below, hence providing a prescription for the
quantization of bottomless actions [20b].

It will be important later to note that in fact $P_Q(\phi)$ is also a
solution of the set of equations
$$0 =(\partial_i S(\phi))P(\phi) + \partial_i P(\phi)\,,\qquad\quad
 i=1,2,\dots
\eqno(2.11)$$
Again one can see from these equations that in the classical limit $P_Q(\phi)$
collapses to a delta function localized at the classical vacuum.

The Langevin equation (2.6) suggests itself a natural definition for
the dynamics
of mixed systems with classical and quantum mechanical degrees of freedom,
namely, to modify the equation by switching off the quantum
fluctuation terms corresponding to the classical fields. More
explicitly, let us consider the set of equations
$$\partial_t\phi_i(t)=-\partial_i S(\phi(t)) + \sqrt{\lambda_i}\eta_i(t)\,,
\qquad\quad i=1,2,\dots \eqno(2.12)$$
where $\lambda_i=0$ for classical degrees of freedom and $\lambda_i=1$
for quantum ones. Note that the same dynamics is obtained by taking
the latter $\lambda_i$ equal to any other common positive number,
instead of unity, since this amounts to a redefinition of the Langevin
time scale. For convenience, we shall keep $\lambda_i$ as free real
and non-negative parameters in the formulae.
Let us stress that here the classical fields are dynamic, not to be confused
with background or external fields which are frozen and
sometimes are also referred to as classical fields in the literature [21].
In the semiquantized system the classical fields do not behave
deterministically, because they acquire stochastic fluctuations through their
coupling to the quantum fields. This was to be expected. A similar phenomenon
occurs in the canonical approach proposals of refs. [9a,9,10],
where the coordinates of the classical particles depend dynamically on
those of the quantum particles and vice versa.

The Fokker-Planck equation is modified to
$$\partial_t P =\partial_i((\partial_i S)P+\lambda_i\partial_i P)\,.
\eqno(2.13)$$
Though the modification introduced by the parameters $\lambda_i$ looks fairly
innocent, in fact they make this equation much more difficult to treat,
even for the stationary solution. Indeed the set of equations, similar
to (2.11),
$$0 =(\partial_i S)P + \lambda_i\partial_i P\,,\qquad\quad
 i=1,2,\dots
\eqno(2.14)$$
is not consistent for generic actions,
unless all the $\lambda_i$ are equal, that is,
the completely classical case ($\lambda_i=0, \forall i$) or
the completely quantized case ($\lambda_i=1, \forall i$). Therefore,
we must look directly for normalizable
solutions of the stationary Fokker-Planck equation:
$$0 =\partial_i((\partial_i S)P+\lambda_i\partial_i P)\,. \eqno(2.15)$$

An important issue in our formulation is the stability of the random walk,
i.e., whether for large enough Langevin times, all normalizable
probability distributions $P(\phi,t)$ evolve to the same normalizable
stationary solution.
In principle, to decide about the stability of the semiquantized system,
one would have to consider the eigenvalue problem corresponding to the
Fokker-Planck eq.~(2.13), and check that there is one non degenerated zero
eigenvalue, whereas all other eigenvalues have a strictly negative real
part. This is not obvious since the construction in eqs.~(2.9,10)
no longer works, and the spectrum may spread over the complex plane.
For arbitrary well-behaved actions, there are at least three types of
possible instabilities, not mutually exclusive:
\item{1.} The spectrum is continuous, hence
there is no normalizable stable distribution (for instance, a Brownian
motion, if the action vanishes). This possibility can be
ruled out, since the action is a strictly increasing polynomial at
infinity and will confine the random walk in the finite configuration space.
\item{2.} There is a zero eigenvalue but it is degenerated.
This is the situation of
the classical dynamics if the action have more than one minimum. Each
of them is a stable fixed point of the Fokker-Planck equation. The
same is true in the semiquantized case if the classical and quantum
sectors are uncoupled. This instability is unlikely to show up if the
classical and quantum sectors are coupled, due to tunneling.
\item{3.} The spectrum contains conjugate pairs of purely imaginary
eigenvalues. This would give rise to stable orbits, rather than points, in
the space of normalizable probability distributions.

We have been able to prove stability of the semiquantized system
if the action is at most quadratic in the fields, as shown in Section 3 below.

In order to compute the vacuum expectation values defined in eq.~(2.1),
we follow the standard procedure of introducing a generating functional
$Z(J)$ [21,15],
$$\eqalign{
Z(J) &= \langle e^{J_i\phi_i}\rangle = \int e^{J_i\phi_i}P(\phi)[d\phi] \cr
&= \sum_{n\ge 0} {1\over n!}\langle\phi_{i_1}\cdots\phi_{i_n}\rangle
J_{i_1}\cdots J_{i_n} \cr }\eqno(2.16)$$
and similarly, a generating functional for the connected vacuum
expectation values, $W(J)$
$$
W(J) = \log Z(J) =
 \sum_{n>0} {1\over n!}\langle\phi_{i_1}\cdots\phi_{i_n}\rangle_c
J_{i_1}\cdots J_{i_n} \eqno(2.17)$$
They are normalized as $Z(0)=1$, and $W(0)=0$.
By taking the Laplace transform of
the stationary Fokker-Planck, eq.~(2.15), one finds
$$ J_i(\lambda_iJ_i - (\partial_i S)({\partial\over\partial J}))Z(J) = 0
\eqno(2.18)
$$
which is a linear differential equation for $Z(J)$, since the action is
polynomial. On the other hand,
the connected generating functional $W(J)$ satisfies a non
linear differential equation.

Before proceeding, let us discuss an important technical point in our
prescription. To quantize a theory with action $S(\phi)/\hbar$ one
can make use of the following generalized Langevin equation (again
in It\^o's sense):
$$\partial_t\phi_i=-{\rm g}^{ij}\partial_j S + \hbar\partial_j {\rm g}^{ij} +
\sqrt{\hbar}v^i_\alpha\eta_\alpha\,,
\qquad\quad i=1,2,\dots \eqno(2.19)$$
where $v^i_\alpha$ can depend on $\phi$ in general and
${\rm g}^{ij}=v^i_\alpha v^j_\alpha$ is known as the kernel of the equation
[22,23,19].
The standard equation corresponds to take $v^i_\alpha=\delta_{i\alpha}$
and thus ${\rm g}^{ij}=\delta_{ij}$.
The generalized equation reproduces the same equilibrium distribution
$P_Q(\phi)$ as the standard one for arbitrary non singular
kernel. This can be seen from the associated Fokker-Planck equation
$$\partial_t P =\partial_i({\rm g}^{ij}((\partial_j S)P
+\hbar\partial_j P)) \eqno(2.20)$$
which also allows for a generalization of the construction in
eqs.~(2.9,10), [19].
The spectrum of the Fokker-Planck Hamiltonian, and hence
the rate to which the random walk thermalizes to its equilibrium
distribution, does depend on the kernel (in fact this one of the
reasons to consider kernelled Langevin equations in practice),
but not the equilibrium distribution itself.
Also the classical limit is independent of the kernel used.
However a semiquantization based on the generalized equation will
depend on the kernel chosen.
Indeed, if we make the replacements $\sqrt{\hbar}\to\sqrt{\lambda_i}$
and $\hbar\to\sqrt{\lambda_i\lambda_j}$ in the Langevin equation, the same
replacement will result in the Fokker-Planck equation. It is immediate
to check that, unless all the $\lambda_i$ coincide, the kernel does not
factor out, even at equilibrium. In other words, we can choose various
inequivalent ways to define the semiquantized theory, all of them
interpolating between the same quantum and classical theories.

It is evident that the standard Langevin equation is not invariant under
changes of coordinates on the configuration manifold R$^N$.
On the other hand, the generalized eq.~(2.19) is covariant under such
transformations, with $v^i_\alpha$ transforming as the contravariant
components of
a tetrad field and ${\rm g}^{ij}$ as the contravariant components of a metric
tensor [19].
In other words, the stochastic quantization formulation requires to
choose a proper Riemannian metric on R$^N$, $ds^2= {\rm g}_{ij}d\phi_id\phi_j$.
The metric is irrelevant at equilibrium in the quantum and classical
theories, but not in the semiquantized theory. Note that the classical
limit, on the other hand, depends on a choice of coordinate system
since the action, and hence its minimum, is not a scalar under
general coordinate transformations.

In what follows we shall consider only the trivial kernel ${\rm
g}^{ij} =\delta_{ij}$.
However there are cases in which a kernel is absolutely needed. For instance,
if the fields $\phi_i$ have different scale dimensions, the standard
metric $ds^2 = d\phi_id\phi_i$, and thus the standard Langevin
equation, violates scale invariance. In the case of relativistic fermion
fields interacting with bosons, a kernel is needed, and in fact there is
a standard choice which reestablishes scale invariance, although it is
introduced for different reasons [11].

\medskip
{\bf 3. Quadratic actions}\par
\medskip
Let us consider an action of the form
$$
S(\phi)=c-h_i\phi_i+{1\over 2}m_{ij}\phi_i\phi_j \eqno(3.1)
$$
with $m_{ij}$ symmetric and strictly positive definite.
The differential equation for the connected generating functional
$W(J)$ (similar to eq.~(2.18)) takes the form
$$J_i(\lambda_i J_i + h_i -m_{ij}\partial_j W(J))=0 \eqno(3.2)$$
where $\partial_j$ refers to partial derivative with respect $J_j$. It
is immediate to check that the most general solution which is analytic at
$J=0$ is a polynomial of second degree in $J_i$
$$W(J)= w_iJ_i + {1\over 2}w_{ij}J_iJ_j \eqno(3.3)$$
By definition of $W(J)$, $w_i=\langle \phi_i \rangle$ and
$w_{ij}=\langle\phi_i\phi_j \rangle_c$. This implies that the dynamics
is Gaussian, since higher order connected expectation values vanish.
Substituting in the equation one finds the conditions
$$
m_{ij}w_j = h_i \,,\quad i=1,\dots, N \eqno(3.4)
$$
$$
\lambda_i\delta_{ij} = {1\over 2}(m_{ik}w_{kj}+w_{ik}m_{kj}),\quad
i,j=1,\dots, N \eqno(3.5)
$$

The first equation expresses that $\langle \phi_i\rangle$ satisfies the
classical equations of motion independently of the choice of
$\lambda_i$. In the quantum case, this is a consequence of Ehrenfest
theorem. The second equation is linear in the the unknowns $w_{ij}$, hence
it can be solved in practice, although we have not found a nice closed
expression for the general solution. Nevertheless, some statements can
be made:
\item{1.} The linear system in eq.~(3.5) is non-singular, since
$m_{ij}$ is strictly positive definite. Hence, the stationary Gaussian
solution (3.3) exists and is unique. Furthermore, the symmetric matrix
$w_{ij}$ is non-negative: let $\{ |n\rangle\,,\quad n=1,\dots,N\}$ be its
complete set of eigenvectors, hence from eq.~(3.5), the eigenvalues
are given by $\langle n|\lambda|n\rangle/\langle n|m|n\rangle$, where
$\lambda$ is the matrix $\lambda_i\delta_{ij}$ and $m$ the matrix
$m_{ij}$, and the $\lambda_i$ non-negative. In fact, $w_{ij}$ is a
positive definite matrix unless there are uncoupled classical degrees
of freedom.
\item{2.} The Gaussian solution (3.3) is in fact
the unique fixed point of the time dependent Fokker-Planck equation.
Indeed, let $\Delta W(\phi,t)$ be the
difference between any other solution and the stationary Gaussian
solution. This difference satisfies the homogeneous equation $(\partial_t +
m_{ij}J_i\partial_j)\Delta W =0$. This equation holds also for the
classical or quantum dynamics, which are stable and hence
$\lim_{t \to +\infty}\Delta W(\phi,t) = 0$.
\item{3.} In the purely classical dynamics all the $\lambda_i$ vanish, and
$w_{ij}$ vanish as well, i.e., there are no fluctuations in the variables
$\phi_i$. Of course, this is true for non quadratic actions too,
since in this case $W(J)=\log Z(J) = w_i J_i$ is the solution of eq.~(2.18).
\item{4.} In the purely quantum dynamics all the $\lambda_i$ equal unity,
and $w_{ij}=(m^{-1})_{ij}$, i.e., $w_{ij}$ is the usual propagator [15].
Comparing with eq.~(3.4), we find the
fluctuation-dissipation theorem, satisfied by the quantum dynamics,
namely, the fluctuation $\langle \phi_i\phi_j\rangle_c$ coincides with
the susceptibility, $\partial\langle \phi_i\rangle/\partial h_j$. This
holds too for generic actions, since
$$
{\partial\langle\phi_i\rangle_J\over\partial J_j} =
 {\partial^2W(J)\over\partial J_i \partial J_j}=
\langle\phi_i\phi_j\rangle_{c,J}
\eqno(3.6)$$
but it is violated by the classical and the semiquantized dynamics.
\item{5.} If the quantum and classical degrees of freedom are
uncoupled, the solution is also straightforward, $w_{ij}$ being block diagonal,
and vanishing in the classical sector. A similar statement
is also true for general actions.
\item{6.} Since the semiquantized dynamics is Gaussian, it describes
a system of non-interacting phonons, as in the quantum case.

As an example, consider a system containing two relativistic fields,
$\phi_{1,2}(x)$, with Euclidean action
$$S(\phi_1,\phi_2)={1\over 2}(\partial\phi_1)^2+{1\over 2}m_1^2\phi_1^2+
{1\over 2}(\partial\phi_2)^2+{1\over 2}m_1^2\phi_2^2+g\phi_1\phi_2
\eqno(3.7)$$
In momentum representation, the mass matrix $m_{ij}$, is
$$m(k)=\left(\matrix{k^2+m_1^2 & g \cr g & k ^2+m_2^2 \cr}\right)
\eqno(3.8)$$
We assume, $m_1^2,m_2^2>0$ and $m_1^2m_2^2>g^2$, so that $m(k)$ is
positive definite.
The equation for the semiquantized connected two point function,
$w_{ij}$, can be solved to give
$$w_{\rm SQ}(k)= {\lambda_2(k^2+m_1^2) +\lambda_1(k^2+m_2^2) \over
(k^2+m_1^2) +(k^2+m_2^2)}w_Q(k) +
{\lambda_1-\lambda_2\over (k^2+m_1^2) +(k^2+m_2^2)}\sigma_3
\eqno(3.9)$$
where $\sigma_3$ is the $z$-component Pauli matrix and
$w_Q(k)$ is the inverse matrix of $m(k)$, i.e., the quantum propagator.
It can be shown explicitly
that $w_{\rm SQ}(k)$ is positive definite, as also follows from the
general argument above. By construction
$$\langle T\phi_i(y)\phi_j(x)\rangle_c^{\rm SQ} =
\int{d^4k\over (2\pi)^4}\exp(-ik(y-x))\,w_{\rm SQ}(k)
\eqno(3.10)$$
The large momentum limit gives
$$w_{\rm SQ}(k)\to {1\over k^2}
\left(\matrix{ \lambda_1 & 0 \cr 0 & \lambda_2 \cr }\right) + {\cal
O}({1\over k ^4})
\eqno(3.11)$$
This implies, from eq.~(3.10),
that
$$\delta(y^0-x^0)\langle [\phi_i(y),\partial_0\phi_j(x)]\rangle =
\lambda_i\delta_{ij}\delta(x-y)
\eqno(3.12)$$
that is, the rescaled equal-time quantum commutation relations. Note that the
disconnected part does not contribute to the commutator.
It is very instructive to study the Lehmann representation. For the
purely quantum case we have
$$w_Q(k) = {P_+ \over k^2+m_+^2} + {P_- \over k^2+m_-^2} \eqno(3.13)$$
where, $m_\pm={1\over 2}(m_1^2+m_2^2 \pm R)$,
$R =  \sqrt{(m_1^2-m_2^2)^2 +4g^2}$,
are the normal masses, and $P_\pm$ are the two orthogonal projectors
onto the normal modes. For the semiquantized case, one finds instead
$$w_{\rm SQ}(k) =
{Q_+ \over k^2+m_+^2} + {Q_- \over k^2+m_-^2}
+ {Q_3\over k^2 + m_3^2}
\eqno(3.14)$$
where, $m_3^2={1\over 2}(m_1^2+m_2^2)$, and
$$\eqalign{
Q_\pm &= \left(
{\lambda_1+\lambda_2\over 2}\pm{(\lambda_1-\lambda_2)(m_1^2-m_2^2) \over 2R}
\right)P_\pm \cr
Q_3 &= {\lambda_1-\lambda_2\over 2}\left(
\sigma_3 -{m_1^2-m_2^2 \over R}(P_+-P_-)\right) \cr
}\eqno(3.15)$$
One can see that there is an extra mode, namely, $m_3^2$.
Unfortunately, whereas
$Q_\pm$ are non negative, $Q_3$ is not, since ${\rm tr}\,(Q_3) = 0$. This
means that the covariance matrix $w_{\rm SQ}(k)$ is positive but
not reflection positive, (except in the trivial cases
$\lambda_1=\lambda_2$ of $g=0$).
As a consequence this theory does not define
a Hilbert space with positive definite metric, i.e., it does
not define a positive physical measure, and for instance, one can
construct operators with negative variance. In other words, the
probabilistic interpretation (of which the classical case is a limit)
breaks down, and the theory must be rejected (or work with a
restricted set of observables, which in this context is ad hoc).
This is a direct consequence of the commutation relations eq.~(3.12).

\medskip
{\bf 4. Feynman rules}\par
\medskip
In order to set up a perturbative calculation let us consider an action
of the form
$$S(\phi) = c - h_i \phi_i +{1\over 2!}m_{ij}\phi_i\phi_j
+ {1\over 3!} g_{ijk}\phi_i\phi_j\phi_k \eqno(4.1)$$
For simplicity, we do not include a quartic term in the action, which
would be required to guarantee stability.
The associated Langevin equation can be brought to the form
$$\phi_l = (\partial_t + m)_{li}^{-1}(\sqrt{\lambda_i}\eta_i +
h_i - {1\over 2!}g_{ijk}\phi_j\phi_k) \eqno(4.2)$$
Because by assumption there is stable equilibrium probability density,
in this equation we can take $(\partial_t+m)^{-1}$ as the retarded
propagator and $\phi_l=0$ at $t\to-\infty$ as boundary condition [11,12].
The equation and its iterative solution is represented in
fig.~1 by means of tree diagrams.
There, the crosses represent quantum fluctuations $\eta_i(t)$ which act
as sources for the fields and the dots are the background sources $h_i$
in the Lagrangian.
Algebraically it is cumbersome but
straightforward to solve the equation iteratively
and compute the $n$-point Euclidean Green function
$\langle\!\langle
\phi_{i_1}(t_1)\phi_{i_2}(t_2)\cdots\phi_{i_n}(t_n)\rangle\!\rangle$
by contracting the $\eta_i(t)$ with the rules
$$
\eqalign{
\langle\!\langle\eta_{i}(t)\eta_{j}(t')\rangle\!\rangle_{\rm c}
 = 2\delta_{ij}\delta(t-t') & \cr
\langle\!\langle
\eta_{i_1}(t_1)\eta_{i_2}(t_2)\cdots\eta_{i_n}(t_n)\rangle\!\rangle_{\rm c}
 = 0 & \qquad\hbox{(\rm for\  $n\not= 2$)} \cr
} \eqno(4.3)
$$
The subindex c stands for the connected part of the expectation value.
Diagrammatically such an expectation value
corresponds to combine $n$ copies of the tree graphs of fig.~1,
by contracting all the crosses pairwise in all possible forms. In this
form the so-called stochastic diagrams are obtained [11,12].
An example is shown
in fig.~2 for the two-point function. (Note that the meaning of the blobs
in both figures is not exactly the same). In general there are several
stochastic diagrams corresponding to each standard Feynman graph, differing
by the positions of the crosses in them. It is noteworthy that if
$\lambda_i=0$, thus removing the quantum fluctuations, only tree level
graphs remain. This corresponds to the classical approximation. Moreover,
because $\lambda_i$ is a kind of $\hbar$ parameter, we find a
selection rule similar to that existing for Feynman graphs [24,21], namely
the number of loops plus the number of external lines in a stochastic
diagram equals the number of crosses plus the number of connected subgraphs.
There is another structural property which is relevant to select the possible
stochastic diagrams: by construction, if the graphs are cut by the crosses
the resulting disconnected pieces are tree graphs containing exactly
one external line.

Actually we only want to compute the $n$-point functions at equilibrium, and
for this purpose it is more convenient to use directly the stationary
Fokker-Planck eq.~(2.15). The probability density itself is quite singular
perturbatively because it is a delta function at zeroth order in the classical
sector, thus we shall use instead the generating functional $Z(J)$, eq.~(2.16).
To simplify, let us remove the external sources, $h_i=0$, and assume a diagonal
representation for the quadratic part, $m_{ij}=s_i^{-1}\delta_{ij}$.
The eq.~(2.18) takes the form
$$0= - s_i^{-1}J_i\partial_iZ - {1\over 2}g_{ijk}J_i\partial_j\partial_kZ
 + \lambda_iJ_i^2Z \eqno(4.4)$$
On the other hand, the connected generated functional $W(J)$,
satisfies the non linear master equation
$$0= s_i^{-1}J_i\partial_iW + {1\over 2}g_{ijk}J_i(
\partial_j\partial_kW + \partial_jW\partial_kW)
 - \lambda_iJ_i^2 \eqno(4.5)$$
In the purely quantum case $\lambda_i=1$, this equation is equivalent
to the set of equations
$$\partial_iW = -{1\over 2}g_{ijk}s_i(
\partial_j\partial_kW + \partial_jW\partial_kW)
 +  s_iJ_i \,, \qquad\quad i=1,2,\dots \eqno(4.6)$$
which is an alternative to the usual construct
$$Z(J) = \exp(-{1\over 3!}g_{ijk}\partial_i\partial_j\partial_k)
\exp({1\over 2}s_lJ_l^2)\eqno(4.7)$$
For unequal $\lambda_i$ only the master equation (4.5) is valid.
Rather than solving this equation perturbatively directly,
it is better to rewrite it as a set of
exact identities among connected Green functions.
Let $w_{i_1 i_2\dots i_n}=\langle\phi_{i_1}\cdots\phi_{i_n}\rangle_{\rm c}$.
Substituting the expansion of $W(J)$ in its master equation, we find
the following hierarchy of identities
$$\eqalign{
w_a     &= -{1\over 2}g_{aij}s_a(w_{ij}+w_iw_j) \cr
w_{ab}  &= -2 s_{ab}[[g_{aij}({1\over 2}w_{bij}+w_jw_{bi})]]
	+ \lambda_as_a\delta_{ab}\cr
w_{abc} &= -3s_{abc}[[g_{aij}({1\over 2}w_{bcij} +
	w_iw_{bcj} + w_{bi}w_{cj})]] \cr
	&\ \vdots \cr
}\eqno(4.8) $$
where $s_{ab\cdots} = (s_a^{-1}+s_b^{-1}+\cdots)^{-1}$, and the symbol
$[[\cdots]]$ means permutation symmetric average on the free indices.
It is worth noting that these identities are exactly the same as in the
quantum theory except the second one, which depends on $\lambda_i$.
These equations can be solved iteratively up to any order in perturbation
theory (we disregard possible non perturbative solutions).
Diagrammatically they are represented in fig.~3 by skeleton equations.
By expanding iteratively these equations, we obtain again the stochastic
diagrams (though they now stand for contributions to vacuum expectation
values, i.e., at equilibrium). Their Feynman rules, which can be read
from eq.~(4.8), are more complicated than
those corresponding to the usual Feynman diagrams. They are the following:
\item{1.} Draw all topologically distinct connected labeled stochastic
graphs and apply the following rules to each one of them.
\item{2.} If the graph is the free two-point graph, apply rule 5.
Otherwise, identify the graph in the skeleton equations in fig.~3.
If this can be done in more than one way, all of them
should be added. Note that the crosses are hidden inside the blobs
and that all but one (uncrossed) line (say, with label $a$)
and one three-point vertex (say, with labels $a,i,j$) will be inside the
blobs.
\item{3.} The value of the graph picks up a factor $s_{abc\cdots}$ from
the average propagator of the external lines. This substitutes the usual
factor $s_a$ in the Feynman rules of standard Feynman graphs.
Also add a factor $-g_{aij}$ from the three-point vertex and a
standard symmetry factor from the structure of the skeleton equation.
\item{4.} Apply rules 2 and 3 to each of the subdiagram which were hidden
in the blobs in previous steps.
\item{5.} The free two-point graph (or subgraph) gives a contribution
$\lambda_a s_a$
\item{6.} Sum over internal indices and add the contributions of different
graphs.

As an illustration, the values of all connected stochastic graphs up to second
order are given in fig.~4.
As a check it can be shown that in the particular case $\lambda_i=1$,
the stochastic graphs add up to the usual Feynman graphs. In the general
case however the average propagators $s_{ijk\cdots}$ remain and this
emphasizes clearly the need for a kernel if the fields have
different dimensions.

It is noteworthy that the semiquantized stochastic graphs do not
satisfy the usual reducibility rules.
This is because the presence of the averaged
propagators $s_{ijk\cdots}$ in the Feynman rules prevents the required
factorization properties to occur. This makes the Feynman rules more
involved than in the purely quantum case and it will
be relevant in the next section
when we discuss the renormalizability of the theory.

\medskip
{\bf 5. Perturbative renormalizability}\par
\medskip
The original interest of studying systems with mixed quantum and classical
degrees of freedom, is the possibility of obtaining finite results in
physically relevant theories which are non renormalizable at the quantum
level, such as gravitation.
The idea is that, given that the classical theory contains no loops and
is finite, a mixed theory would be more ultraviolet convergent than the
completely quantum case.

In order to clarify this issue we can study a Lagrangian such as
$${\cal L}(x) = {\cal L}_{\rm KG}(\phi) + {\cal L}_{\rm KG}(\psi) +
{1\over 2}g\phi(x)\psi^2(x) + {1\over 4!} \gamma\phi^4(x) \eqno(5.1)$$
${\cal L}_{\rm KG}$ being the Klein-Gordon Lagrangian. In six space-time
dimensions this theory is not renormalizable unless the parameter $\gamma$
vanishes [26].
The interesting semiquantized case is when $\psi(x)$ is a
quantum field and $\phi(x)$ is classical. The other way around gives a
trivial theory with no loops. For our discussion it will be enough
to consider the case $\gamma=0$ in six dimensions.
This semiquantized theory turns out to be non renormalizable. To see this,
consider the stochastic
graph in fig.~$5a$. It is the lowest order divergent graph for the two-point
Green function of the field $\phi(x)$. In the quantum theory such a divergence
would be renormalized by redefining the mass and wavefunction of the field.
The counterterm is shown in fig.~$5b$. In the semiquantized theory, however,
there is no such contribution because $\lambda_\phi$ vanishes. Note that
other second order stochastic graphs analogous to that in fig.~$5a$ but
differing in the position of the crosses also vanishes for the same reason.
Similarly, the lowest order divergent vertex correction, shown in fig.~$6a$,
cannot be canceled by the usual counterterm, fig.~$6b$, which does not exists
in the semiquantized theory. The conclusion is that the semiquantized
version of the original renormalizable quantum theory is non renormalizable,
at least perturbatively. Clearly
the same conclusion will hold for more complicated theories too (for instance
letting $\gamma\not=0$).

A possible way out would be to take $\lambda_\phi$ as a free parameter with
the prescription $\lambda_\phi\to 0^+$ at the end. This would allow a mass
and wavefunction
counterterm of order $g^2\lambda_\phi^{-1}$ to cancel the mass and wavefunction
correction divergent graph, and similarly for the vertex correction.
The analysis
of this procedure at higher orders is very involved. In particular
note that in the quantum theory, reducible graphs are automatically
renormalized once irreducible graphs are, but this does not hold in the
semiquantized case. This implies that for instance the two bubble graph
in fig.~7 requires a further mass and wavefunction counterterm of
order $g^4\lambda_\phi^{-2}$. On the other hand it is not clear
whether such a prescription is effectively cancelling the limit
$\lambda_\phi\to 0$ and thus going over to the full quantum theory.

The problem with the renormalizability is closely related to the fact that
the stochastic diagrams lack good reducibility properties and hence it is
more
a problem of the stochastic approach than of the semiquantization itself.
Indeed by taking arbitrary $\lambda_i$ we are weighting differently the
various stochastic graphs. What we see is that they are not renormalizable
independently. Likely, this implies that even in the standard quantum
case, the expectation values are not
finite until the random walk reaches its equilibrium distribution.

\medskip
{\bf 6. Effective action}\par
\medskip

In the quantum case, the variables $J_i$ are identified with currents
in the sense that they create excited states on the vacuum. However, this
is not true classically. For instance, if the classical Euclidean
vacuum corresponds to $\phi_i=0$, $Z(J)=\langle e^{J_i\phi_i}\rangle$
is identically equal to unity, since there are no fluctuations.
The correct way to introduce a current $h_i$
to pick up excited states, valid both in the quantum and classical
cases, is through the action, i.e., by considering the
family of actions $S(\phi)-h_i\phi_i$. Hence one must consider rather
$Z(J,h)=\langle e^{J_i\phi_i}\rangle_h$ as the generator of
expectations values in the presence of external currents; in the
quantum case $Z(J,h)$ depends on $J_i+h_i$ only. In order to define
the effective action, let $\phi_i(h)=\langle\phi_i\rangle_h$ be the
so-called classical field, and let $\Omega(h)$ its connected
generator, i.e.,
$$ \phi_i(h)={\partial\Omega(h)\over\partial h_i } \eqno(6.1)$$
by definition. First, one must show that $\Omega(h)$ exists, that is,
that the integrability conditions of eq.~(6.1) are met. This is true
in the quantum case ($\Omega$ is just $W$), and in the classical case:
$\Omega$ is the Legendre transform of the action. The proof follows
from the Langevin equation,
$$\partial_t\phi_i(t,h)=-\partial_i S+h_i-\sqrt{\lambda_i}\eta_i\eqno(6.2)$$
Applying $\partial/\partial h_j$ one finds
$${\partial\phi_i(t,h)\over\partial h_j}=(\partial_t+\partial^2S)^{-1}_{ij}
\eqno(6.3)$$
independently of $\lambda_i$. Since the right hand side is manifestly
symmetric in the indices $ij$, this is also true after average over
$\eta_i$ and in the limit of large Langevin time, i.e., the matrix
${\partial\phi_i(h)/\partial h_j}$ is also symmetric and $\Omega$
exists in the semiquantized case. Note that in general
${\partial\phi_i(h)/\partial h_j}$ does not coincide with
$\langle\phi_i\phi_j\rangle_{h,c}$. They coincide in the quantum case,
but in the classical case the latter expression vanishes.
The effective action is defined by the Legendre transform of $\Omega(h)$,
$$\Gamma(\phi) = -\Omega(h)+h_i\phi_i, \eqno(6.4)$$
This definition is the usual effective action in the quantum case and
is the action $S(\phi)$ in classical case.
The best way to compute it is by solving
$$h_i={\partial\Gamma(\phi)\over\partial\phi_i}\eqno(6.5)$$
For the action $S(\phi)={1\over 2}s_i^{-1}\phi_i^2
+{1\over 3!}g_{ijk}\phi_i\phi_j\phi_k$, one finds perturbatively
$$\Gamma(\phi)={1\over 2}s_i^{-1}\phi_i^2 +{1\over 3!}g_{ijk}\phi_i\phi_j\phi_k
+{1\over 2}g_{ijj}s_j\lambda_j\phi_i -
{1\over 2}g_{ijk}g_{\ell jk}s_{jk}s_j\lambda_j\phi_i\phi_\ell
+{\cal O}(g^3)\eqno(6.6)$$
corresponding respectively to the diagrams in fig.~8. It is
interesting to note that the effective action only contains one particle
irreducible graphs and it is regular as $\lambda_i\to 0$. This is
unlike the Legendre transform of $W(J)$. It implies that the effective
action is renormalizable if the action is renormalizable. However,
recall that the effective action only generates the expectation value
of $\phi_i$, not the two-point function, etc.

\medskip
{\bf 7. Conclusions}\par
\medskip
We have defined and
studied some aspects of the dynamics of semiquantized fields.
These systems interpolate between the completely quantum and the
completely classical theories. The semiquantization is not uniquely
defined, however. As pointed out above the choice of a concrete kernel
is also needed in the stochastic quantization itself, and it is the origin
of quantum anomalies in this formulation.
The study shows that this kind of theories present two flaws. First,
they break positivity: they do not define a positive measure for the
Minkowskian expectation values. The other conflict comes
from their renormalizability. Here we found a surprise, namely,
the removal of quantum
fluctuations in a subset of the degrees of freedom does not necessarily
make the theory more ultraviolet convergent. On the contrary, in general
(and in fact in the interesting cases), the
renormalizability is spoiled by the semiquantization.
This is because in the semiquantized
theory we are setting to zero a subset of (stochastic) Feynman graphs
and the remaining diagrams no longer form a closed set under renormalization.
There are not enough counterterms left to cancel all the divergences.
Less technically, the lesson from fig.~5 or 6 is that the missing
intrinsic fluctuations of the classical field are needed to compensate
the fluctuations induced through the coupling to the quantum field.
In fact the problem with positivity, renormalizability, as well as the lack
of uniqueness and of good reducibility properties of the Green functions,
is a manifestation of the fact that the equations of the semiquantized
theory are considerably less symmetric than those of the quantum or
classical theories. We think that this kind of problems will appear also
in the other approaches to semiquantization existing in the literature,
and in our view this means that the concept of semiquantization is
rather unnatural.

\medskip
{\bf Acknowledgments}\par
\medskip
This work has been partially supported by DGICYT contract no. PB92-0927.
I thank C. Garc\'{\i}a-Recio for reading the manuscript.
\vfill\eject
\null\medskip
{\bf References}\par
\medskip
\item{[1]}S. Hawking and W. Israel, General Relativity,(Cambridge, 1979).
\item{[2]} P. Van Nieuwenhuizen, Phys. Reports 68 (1981) 189.
\item{[3]} S. Adler, Rev. Mod. Phys. 54 (1982) 729.
\item{[4]} M.B. Green, J.H. Schwarz and E. Witten,
	``Superstring Theory'', (Cambridge, 1987).
\item{[5]} C. M{\o}ller, in Les Theories relativistes de la
	gravitacion, CNRS, Paris 1962.
\item{[6]} V.N. Lukash, I.D. Novikov, A.A. Starobinsky and
	Ya.B. Zel'dovich, Nuovo Cimento 35B (1976) 293.
\item{[7]} T.W.B. Kibble, in Quantum Gravity 2, C.J. Isham,
	R. Penrose and D.W.Sciama eds. (Clarendon, 1981).
\item{[8]} B.L. Hu and A. Matacz, Univ. of Maryland preprint, umdpp 94-31.
\item{[9a]} I.V. Aleksandrov, Z. Naturforsch. 36A (1981) 902.
\item{[9]} W. Boucher and J. Traschen, Phys. Rev. D37 (1988) 3522.
\item{[10]} A. Anderson, Phys. Rev. Lett. 74 (1995) 621.
\item{[10a]} L. Di\'osi, {\it Comment on `Quantum Backreaction on
``Classical''Variables'}, submitted to Phys. Rev. Lett.
\item{[10b]} A. Anderson, {\it Reply to Comment on `Quantum Backreaction on
``Classical''Variables'}, submitted to Phys. Rev. Lett.
\item{[11]} P.H. Damgaard and H. H\"uffel,
	Phys. Reports 152 (1987) 227.
\item{[12]} M. Namiki, I. Ohba, K. Okano, Y. Yamanaka, A.K. Kapoor,
        H. Nakazato and S. Tanaka, ``Stochastic quantization'',
       (Springer, Lecture notes in physics, Berlin, 1992).
\item{[13]} G. Parisi and Y.-S. Wu, Sci. Sinica 24 (1981) 483.
\item{[14]} J. Glimm and A. Jaffe, ``Quantum Physics'' (Springer Verlag,
	1987).
\item{[15]} P. Ramond, ``Field theory: a modern primer'', (Addison Wesley,
	1990).
\item{[16]} G. Roepstorff, ``Path Integral Approach to Quantum Physics'',
	(Springer Verlag, 1994).
\item{[17]} G. Parisi, Phys. Lett. B131 (1983) 393.
\item{[18]} J.R. Klauder, Phys. Rev. A29 (1984) 2036.
\item{[19]} L.L. Salcedo, Phys. Lett. B304 (1993) 125.
\item{[20]} G.G. Bartrouni, G.R. Katz, A.S. Kronfeld, G.P. Lepage,
	B. Svetisky and K.G. Wilson, Phys. Rev. D32 (1985) 2736.
\item{[20a]} E. Gozzi, Phys. Lett. 150B (1985) 119.
\item{[20b]} J. Greensite and M.B. Halpern, Nucl. Phys. B242 (1984) 167;
        J. Greensite, Nucl. Phys. B439 (1993) 439.
\item{[21]} C. Itzykson and J.-B. Zuber, ``Quantum Field Theory'',
	(McGraw-Hill, 1980).
\item{[22]} J.D. Breit, S. Gupta and A. Zaks, Nucl. Phys. B233 (1984) 61.
\item{[23]} P.H. Damgaard and K. Tsokos, Nucl. Phys. B235 (1984) 75.
\item{[24]} S. Coleman and E. Weinberg, Phys. Rev. D7 (1973) 1888.
\item{[25]} M. Creutz, ``Quarks, gluons and lattices'', (Cambridge, 1983).
\item{[26]} J. Collins, ``Renormalization'' (Cambridge, 1984).
\vfill\eject
\null\medskip
{\bf Figure captions}\par
\medskip
\noindent
Fig.~1: Diagrammatic representation of the Langevin equation (4.2)
	and its iterative solution.
\par\noindent
Fig.~2: Typical stochastic graph for the two-point function.
\noindent
Fig.~3: Diagrammatic representation of the hierarchy of identities
	in eq.~(4.8)
\par\noindent
Fig.~4: Connected stochastic graphs and their values up to second order
	in perturbation theory.
\par\noindent
Fig.~5: Two-point divergent stochastic graph at lowest order (a) and
	its counterterm graph (b).
\par\noindent
Fig.~6: Three-point divergent stochastic graph at lowest order (a) and
	its counterterm graph (b).
\par\noindent
Fig.~7: Divergent reducible two loop graph.
\par\noindent
Fig.~8: Effective action graphs up to second order.

\vfill\eject

\bye

%
\input texdraw
\def\blob{\bsegment \lcir r:0.5 \fcir f:0.6 r:0.5 \esegment}
\def\dot{\bsegment \fcir f:0 r:0.08 \esegment}
\def\sdot{\rmove(0.15 0)\bsegment \fcir f:0 r:0.03 \esegment\rmove(0.15 0)}
\def\cross{\bsegment\rmove(0.1 -0.1)\rlvec(-0.2 0.2)\rmove(0.2 0)
	\rlvec(-0.2 -0.2)\esegment}
\def\vcross{\bsegment\rmove(0 0.15)\rlvec(0 -0.3)\rmove(-0.15 0.15)
	\rlvec(0.3 0)\esegment}
\def\igual{\bsegment\move(0 0.08)\rlvec(0.4 0)\move(0 -0.08)\rlvec(0.4 0)
	\esegment}
\def\mas{\bsegment\rlvec(0.4 0)\move(0.2 0.2)\rlvec(0 -0.4)\esegment
	\rmove(0.4 0)}
\def\Dpie #1{\bsegment\textref h:C v:C \htext(0 -0.75){#1}\esegment}
\def\dpie #1{\bsegment\textref h:C v:C \htext(0 -0.5){#1}\esegment}
\def\upie #1{\bsegment\textref h:C v:C \htext(0 +0.5){#1}\esegment}
\def\rpie #1{\bsegment\textref h:C v:C \htext(+0.5 0){#1}\esegment}

\btexdraw
\drawdim cm \setunitscale 0.8 \linewd 0.03
\move(-1 0)
\rlvec(1.5 0) \rmove(0.5 0) \blob\rmove(0 -0.25)\Dpie{$\phi$}\rmove(0 0.25)
\rmove(1.25 0) \savecurrpos(*x1 *y1)\igual\rmove(1.25 0)
\rlvec(1.5 0)\cross\Dpie{$\eta$}
\rmove(0.75 0)\mas\rmove(0.75 0)\rlvec(1.5 0)\dot\Dpie{$h$}
\rmove(0.75 0)\mas\rmove(0.75 0)\rlvec(1.5 0)
	\savecurrpos(*x *y)\rlvec(1 1)\rmove(0.3535 0.3535)
	\blob\move(*x *y)\rlvec(1 -1)\rmove(0.3535 -0.3535)\blob
\move(*x1 *y1)\rmove(0 -4)\igual\rmove(1.25 0)
	\sdot\sdot\sdot\rmove(0.75 0)\mas\rmove(0.75 0)\rlvec(1.5 0)
	\savecurrpos(*x *y)\rlvec(1 1)\dot\rmove(-1 -1)\rlvec(1 -1)
	\rlvec(1.5 0)\cross\rmove(-1.5 0)\rlvec(0 -1.5)\cross
\move(*x *y)\rmove(3 0)\rmove(0.75 0)\mas\rmove(0.75 0) \sdot\sdot\sdot
\etexdraw
\vskip 24pt
\centerline{Fig.~1.}
\vskip 4cm
\btexdraw
\drawdim cm \setunitscale 0.8 \linewd 0.03
\lvec(1 0)\rmove(0.5 0)\blob\rmove(0.5 0)\rlvec(1 0)\rmove(1 0)\igual
	\rmove(1 0)\sdot\sdot\sdot\rmove(0.5 0)\mas\rmove(0.5 0)
\rlvec(1 0)\rlvec(0 -1)\dot\rmove(0 1)\rlvec(1 0)\rmove(1 0)
	\lcir r:1 \rmove(0 1)\cross\rmove(1 -1)\rlvec(1 0)\cross\rlvec(1 0)
	\rmove(0.5 0)\mas\rmove(0.5 0)\sdot\sdot\sdot
\etexdraw
\vskip 24pt
\centerline{Fig.~2.}

\vfill\eject
\btexdraw
\drawdim cm \setunitscale 0.7 \linewd 0.03
\move(-1 0)
\bsegment 
\rmove(0.25 0)\rlvec(1.5 0)\rmove(0.5 0)\blob\rmove(0.5 0)\rmove(0.25 0)
	\rmove(1 0)\igual\savecurrpos(*igualx *igualy)\rmove(1 0)
	\rlvec(1 0)
	\rmove(0.653 0)\larc r:0.653 sd:45 ed:315 \rmove(0.635 0)\blob
	\rmove(0.5 0)\rmove(1 0)
	\rmove(0.5 0)\savecurrpos(*mx1 *my1)\mas\rmove(0.5 0)
	\rlvec(1.5 0)
	\savecurrpos(*x *y)\rlvec(1 1)\rmove(0.3535 0.3535)
	\blob\move(*x *y)\rlvec(1 -1)\rmove(0.3535 -0.3535)\blob
\esegment
\rmove(0 -5)
\bsegment 
\rlvec(1 0)\rmove(0.5 0)\blob\rmove(0.5 0)\rlvec(1 0)
	\move(*igualx *igualy)\rmove(0 -5)\igual\rmove(1 0)
	\rlvec(1 0)
	\rmove(0.653 0)\larc r:0.653 sd:45 ed:315 \rmove(0.635 0)\blob
	\rmove(0.5 0)\rlvec(1 0)
	\move(*mx1 *my1)\rmove(0 -5)\mas\rmove(0.5 0)
	\rlvec(1.5 0)
	\savecurrpos(*x *y)\rlvec(1 1)\rmove(0.3535 0.3535)
	\blob\move(*x *y)\rlvec(1 -1)\rmove(0.3535 -0.3535)\blob
	\rmove(0.5 0)\rlvec(1 0)
	\move(*x *y)\rmove(3 0)
	\rmove(0.5 0)\savecurrpos(*mx2 *my2)\mas\rmove(0.5 0)
	\rlvec(1.5 0)\cross\rlvec(1.5 0)
\esegment
\rmove(0 -5)
\bsegment 
\rlvec(1 0)\rmove(0.5 0)\blob\savecurrpos(*x *y)\rmove(0.3535 0.3535)
	\rlvec(1 1)\move(*x *y)\rmove(0.3535 -0.3535)\rlvec(1 -1)
	\move(*igualx *igualy)\rmove(0 -10)\igual\rmove(1 0)
	\rlvec(1 0)
	\rmove(0.653 0)\larc r:0.653 sd:45 ed:315 \rmove(0.635 0)\blob
	\savecurrpos(*x *y)\rmove(0.3535 0.3535)
	\rlvec(1 1)\move(*x *y)\rmove(0.3535 -0.3535)\rlvec(1 -1)
	\move(*mx1 *my1)\rmove(0 -10)\mas\rmove(0.5 0)
	\rlvec(1.5 0)
	\savecurrpos(*x *y)\rlvec(1 1)\rmove(0.3535 0.3535)
	\blob\move(*x *y)\rlvec(1 -1)\rmove(0.3535 -0.3535)\blob
	\savecurrpos(*x *y)\rmove(0.3535 0.3535)
	\rlvec(1 1)\move(*x *y)\rmove(0.3535 -0.3535)\rlvec(1 -1)
	\move(*x *y)\rmove(3 0)
	\move(*mx2 *my2)\rmove(0 -5)\mas\rmove(0.5 0)
	\rlvec(1.5 0)
	\savecurrpos(*x *y)\rlvec(1 1)\rmove(0.3535 0.3535)
	\blob\rmove(0.5 0)\rlvec(1 0)
	\move(*x *y)\rlvec(1 -1)\rmove(0.3535 -0.3535)\blob
	\rmove(0.5 0)\rlvec(1 0)
\esegment
\etexdraw
\vskip 24pt
\centerline{Fig.~3.}
\vfill\eject
\btexdraw
\drawdim cm \setunitscale 0.8 \linewd 0.03
\bsegment 
\rmove(0.5 0)\dpie{$a$}\rlvec(1.5 0)\cross\rlvec(1.5 0)\dpie{$b$}\rmove(0.5 0)
\rmove(2 0)
\textref h:L v:C \htext{
	$s_a\delta_{ab}\lambda_a$}
\esegment
\rmove(0 -3)
\bsegment 
\rmove(0.5 0)\dpie{$a$}\rlvec(1 0)\rmove(1 0)\lcir r:1
	\rmove(1 0)\cross\Dpie{$i$}\rmove(0.5 0)
\rmove(2 0)
\textref h:L v:C \htext{
	$-{1\over 2}g_{aii}s_as_i\lambda_i$}
\esegment
\rmove(0 -3)
\bsegment 
\rmove(0.75 0)\dpie{$a$}\rlvec(1.5 0)\rlvec(0.5 0.5)\vcross\rpie{$b$}
	\rlvec(0.5 0.5)
	\rmove(-1 -1)\rlvec(0.5 -0.5)\vcross\rpie{$c$}\rlvec(0.5 -0.5)
	\rmove(0 1)\rmove(0.75 0)
\rmove(2 0)
\textref h:L v:C \htext{
	$-g_{abc}s_{abc}s_bs_c\lambda_b\lambda_c$}
\esegment
\rmove(0 -3)
\bsegment 
\rmove(1 0)\rlvec(-0.5 0.5)\vcross\rlvec(-0.5 0.5)\dpie{$b$}\rmove(1 -1)
	\rlvec(-1 -1)\upie{$a$}\rmove(1 1)\rlvec(0.5 0)\dpie{$i$}\rlvec(0.5 0)
	\rmove(1 0)\lcir r:1 \rmove(1 0)\cross\Dpie{$j$}
	\rmove(2 0)
	\textref h:L v:C \htext{
	 ${1\over 2}g_{abi}g_{ijj}s_{ab}s_bs_is_j
	\lambda_b\lambda_j$}
\esegment
\rmove(0 -3)
\bsegment 
\dpie{$a$}\rlvec(0.5 0)\cross\rlvec(0.5 0)\rmove(1 0)\lcir r:1 \rmove(0 1)
	\cross\dpie{$i$}\rmove(0 -2)\upie{$j$}\rmove(1 1)\rlvec(1 0)\dpie{$b$}
\rmove(2 0)
\textref h:L v:C \htext{
	$g_{aij}g_{bij}s_as_is_{aij}s_{ab}\lambda_a\lambda_i$}
\esegment
\rmove(0 -3)
\bsegment 
\dpie{$a$}\rlvec(0.5 0)\rlvec(0.5 0)\rmove(1 0)\lcir r:1 \rmove(0 1)\cross
	\dpie{$i$}\rmove(0 -2)\cross\upie{$j$}\rmove(1 1)\rlvec(1 0)\dpie{$b$}
\rmove(2 0)
\textref h:L v:C \htext{
	${1\over 2}g_{aij}g_{bij}s_{ab}s_is_j
	(s_{aij}+s_{bij})\lambda_i\lambda_j$}
\esegment
\rmove(0 -3)
\bsegment 
\rmove(1 0)\rlvec(-0.5 0.5)\vcross\rlvec(-0.5 0.5)\dpie{$b$}\rmove(1 -1)
	\rlvec(-1 -1)\upie{$a$}\rmove(1 1)\rlvec(1 0)\dpie{$i$}\rlvec(1 0)
	\rlvec(0.5 0.5)\vcross\rlvec(0.5 0.5)\dpie{$c$}\rmove(-1 -1)
	\rlvec(0.5 -0.5)\vcross\rlvec(0.5 -0.5)\upie{$d$}\rmove(0 1)
\rmove(2 0)
\textref h:L v:C \htext{
	$g_{abi}g_{cdi}s_{abcd}s_bs_{cdi}s_cs_d
	\lambda_b\lambda_c\lambda_d$}
\esegment
\rmove(0 -3)
\bsegment 
\rmove(1 0)\rlvec(-0.5 0.5)\vcross\rlvec(-0.5 0.5)\dpie{$b$}\rmove(1 -1)
	\rlvec(-1 -1)\upie{$a$}\rmove(1 1)\rlvec(1 0)\cross\dpie{$i$}
	\rlvec(1 0)
	\rlvec(0.5 0.5)\vcross\rlvec(0.5 0.5)\dpie{$c$}\rmove(-1 -1)
	\rlvec(0.5 -0.5)\rlvec(0.5 -0.5)\upie{$d$}\rmove(0 1)
\rmove(2 0)
\textref h:L v:C \htext{
	$g_{abi}g_{cdi}s_{abcd}s_bs_is_c
	(s_{abi}+s_{cdi})\lambda_b\lambda_c\lambda_i$}
\esegment
\etexdraw
\centerline{Fig.~4.}
\vfill\eject
\btexdraw
\drawdim cm \setunitscale 1.2 \linewd 0.02
\bsegment 
\dpie{$\phi$}\rlvec(1 0)\rmove(1 0)\lpatt(0.067 0.1)\lcir r:1
	\rmove(0 -1.5)\dpie{$(a)$}\rmove(0 1.5)
	\lpatt()\rmove(0 1)\cross
	\rmove(0 -2)\cross\rmove(0.3 0)\rpie{$\psi$}\rmove(-0.3 0)
	\rmove(1 1)\rlvec(1 0)
\rmove(2 0)
\rlvec(0.75 0)\cross\rlvec(0.75 0)\dot
	\rmove(0 -1.5)\dpie{$(b)$}\rmove(0 1.5)
	\rlvec(1.5 0)
\esegment
\etexdraw
\vskip 24pt\centerline{Fig.~5.}
\vskip 3cm
\btexdraw
\drawdim cm \setunitscale 1.2 \linewd 0.02
\bsegment 
\lpatt(0.067 0.1)\lcir r:1 \lpatt()
	\rmove(1 0)\cross\move(0 0)
	\rmove(-0.5 0.866)\cross\move(0 0)
	\rmove(-0.5 -0.866)\cross\move(0 0)
	\rmove(-1 0)\rlvec(-1 0)\move(0 0)
	\rmove(0.5 0.866)\rlvec(0.5 0.866)\move(0 0)
	\rmove(0.5 -0.866)\rlvec(0.5 -0.866)\move(0 0)
	\rmove(0 -2)\dpie{$(a)$}\move(0 0)
\move(5 0)\savecurrpos(*x *y)
	\dot
	\rlvec(-1 0)\move(*x *y)
	\rlvec(0.25 0.433)\vcross\rlvec(0.25 0.433)\move(*x *y)
	\rlvec(0.25 -0.433)\vcross\rlvec(0.25 -0.433)\move(*x *y)
	\rmove(0 -2)\dpie{$(b)$}\move(*x *y)
\esegment
\etexdraw
\vskip 24pt\centerline{Fig.~6.}
\vfill\eject
\btexdraw
\drawdim cm \setunitscale 1.2 \linewd 0.02
\bsegment 
\rlvec(1.5 0)\rmove(1 0)\lpatt(0.067 0.1)\lcir r:1 \lpatt()
	\rmove(0 1)\cross\rmove(0 -2)\cross\rmove(1 1)
\rlvec(1.5 0)\rmove(1 0)\lpatt(0.067 0.1)\lcir r:1 \lpatt()
	\rmove(0 1)\cross\rmove(1 -1)\rlvec(1.5 0)
\esegment
\etexdraw
\vskip 24pt\centerline{Fig.~7.}
\vskip 2cm
\btexdraw
\drawdim cm \setunitscale 0.8 \linewd 0.03
\bsegment 
\rmove(0.5 0)\rlvec(3. 0)\rmove(0.5 0)
\rmove(0.5 0)\mas\rmove(0.5 0)
\rmove(0.75 0)\rlvec(1.5 0)\rlvec(1. 1.)\rmove(-1 -1)\rlvec(1. -1.)
	\rmove(0 1)\rmove(0.75 0)
\esegment
\rmove(0 -3)
\rmove(0.5 0)\mas\rmove(0.5 0)
\bsegment 
\rmove(0.5 0)\rlvec(1 0)\rmove(1 0)\lcir r:1
	\rmove(1 0)\cross\rmove(0.5 0)
\esegment
\rmove(4 0)\mas\rmove(0.5 0)
\bsegment 
\rlvec(0.5 0)\rlvec(0.5 0)\rmove(1 0)\lcir r:1 \rmove(0 1)\cross
	\rmove(0 -2)\rmove(1 1)\rlvec(1 0)
\esegment
\etexdraw
\vskip 24pt
\centerline{Fig.~8.}

\bye